\documentclass[a4paper,12pt]{article}
\usepackage{amsmath,amssymb,amsthm}
\usepackage{latexsym,graphicx,color,bbm,subfigure}
\usepackage{pdfsync}

\newcommand{\beq}{\begin{eqnarray}}
\newcommand{\eeq}{\end{eqnarray}}

\newcommand{\Tr}{{\rm Tr}}
\newcommand{\diag}{{\rm diag}}

\newcommand{\bea}{\begin{eqnarray}}
\newcommand{\eea}{\end{eqnarray}}

\def\de{\partial}
\def\im{\hbox{\rm Im}}

\def\bra{\langle}
\def\ket{\rangle}

\begin{document}

\thispagestyle{empty}
\begin{flushright}
IFUP-TH/2010-21
\end{flushright}
\vspace{10mm}
\begin{center}
{\Large \bf Monopole-vortex complex  in a $\theta$  vacuum
} 
\\[15mm]
{Kenichi~{\sc Konishi}}$^{a,b}$ \footnote{\it e-mail address:
konishi(at)df.unipi.it}
{Alberto~{\sc Michelini}}$^{c,b}$ \footnote{\it e-mail address:
a.michelini(at)sns.it},
{Keisuke~{\sc Ohashi}}$^{d}$ \footnote{\it e-mail address:
ohashi(at)gauge.scphys.kyoto-u.ac.jp}
\vskip 6 mm

\bigskip\bigskip
{\it
%$^1$  
~Department of Physics ``E. Fermi'', University of Pisa  $^{a}$, \\
Largo Pontecorvo, 3, Ed. C, 56127 Pisa, Italy
\\
%$^2$
  INFN, Sezione di Pisa $^{b}$,
Largo Pontecorvo, 3, Ed. C, 56127 Pisa, Italy 
\\
%$^2$
  Scuola Normale Superiore $^{c}$, 
Piazza dei Cavalieri, 7, Pisa, Italy  
\\
%$^2$
  Department of Physics, Kyoto  University $^{d}$, 
Kyoto 606-8502, Japan
 }

\vskip 6 mm

\bigskip
\bigskip

{\bf Abstract}\\[5mm]
{\parbox{14cm}{\hspace{5mm}
\small
%\preprint{IFUP-TH/2010-21}

We discuss aspects of the monopole-vortex complex soliton   arising in a hierarchically broken gauge system, $G \to H \to {\mathbbm 1} $, in a $\theta$ vacuum
of the underlying $G$ theory. Here we focus our attention mainly on the simplest such system with  $G=SU(2)$ and $H=U(1)$.   A consistent picture of the effect of the $\theta$ parameter is found both in a macroscopic, dual picture and in a microscopic description of the monopole-vortex complex soliton.

}}
\end{center}
\newpage
\pagenumbering{arabic}
\setcounter{page}{1}
\setcounter{footnote}{0}
\renewcommand{\thefootnote}{\arabic{footnote}}

\section{Introduction}

 It was suggested recently \cite{ABEKY}-\cite{Konishi} that certain properties of the regular 't Hooft-Polyakov monopoles \cite{TH} arising from a gauge symmetry
 breaking $G \longrightarrow  H$   occurring at some mass scale $v_{1}$  may be best studied by putting the low-energy $H$ system in a  Higgs phase,  by the  vacuum expectation values (VEV)   $v_{2}$  of another scalar field (see also \cite{CHA2}), so that one has a hierarchical gauge symmetry breaking pattern, 
\[
G \stackrel{v_{1}}{\longrightarrow}   H  \stackrel{v_{2}}{\longrightarrow}   {\mathbbm 1}\;,  \qquad  v_{1}  \gg   v_{2}\;.  
\]
   The regular  monopoles correspond to the second homotopy group $\pi_{2}(G/H)$ while the vortex solutions describe nontrivial elements of the fundamental group $  \pi_{1}(H)$.   They are related by the exact  homotopy-group sequence,
\[    \cdots \to   \pi_{2}(G)   \to    \pi_{2}(G/H)   \to      \pi_{1}(H)     \to   \pi_{1} (G) \to \cdots \,.   
\] 
By using the known fact that $\pi_{2}(G) = {\mathbbm 1}$ for any compact Lie group $G$,  one finds that 
\[     \pi_{1}(G)    \sim      \pi_{1}(H) / \pi_{2}(G/H)   \;.
\]
For instance if  the original gauge group is simply connected  (e.g., $G=SU(N)$)
  there is  one-to-one correspondence between a vortex solution and a regular monopole solution.   Physically this means that the latter
 disappears from the spectrum, as it is confined by the vortex of the low-energy system. Vice versa, the vortex of 
low-energy system is unstable in the sense that it terminates at massive monopoles at its extremes (or equivalently, can be cut in the middle by a pair production of massive monopoles).  If   $\pi_{1}(G) \ne {\mathbbm 1}$ this relation is generalized appropriately. For instance for $G=SO(N)$,  $\pi_{1}(G) ={\mathbbm Z}_{2}$, the correspondence is two-to one:   doubly-wound vortices are unstable and require a regular monopole to exist in the system \cite{TH}\;.  

This kind of relation gives powerful information on the monopoles when the vortex properties are known, or vice versa. 
For instance, when the low-energy vortices carry continuous, non-Abelian orientational moduli such as those studied extensively recently \cite{HT,ABEKY}, \cite{SY}-\cite{GJK}, consistency requires the massive monopoles at the vortex extremities to possess corresponding non-Abelian zeromodes. This seems to explain the occurrence  of fully quantum mechanical  non-Abelian monopoles in the infrared spectrum of certain ${\cal N} =2$  supersymmetric quantum chromodynamics (QCD) \cite{APS}-\cite{CKM}. 

The very nature of the problem thus  leads us to the study of soliton complexes which are not stable.  When there is a strong  hierarchy of the symmetry breaking scales,  $v_{1} \gg v_{2}$, however, we can appeal to  a Born-Oppenheimer type approximation:  the motion of the heavy monopole can be neglected in the analysis 
of low-energy excitations of the vortex-monopole complex. 

An important point of this kind of analysis is that certain relations between the monopoles and vortices,  following from symmetry, consistency and continuity, must hold  exacty.  The first check of such a connection (the Abelian {\it and}  non-Abelian magnetic flux matching) has been made  in \cite{ABEK}, soon after the discovery of non-Abelian vortex solutions. 

It is the purpose of this note to examine the properties of a monopole-vortex complex in a $\theta$ vacuum.  We wish to know how the $\theta$ parameter  affects the whole system.

In order to study the question in the simplest possible context, we mainly discuss below the case of symmetry breaking,
\beq
SU(2) \stackrel{v_{1}}{\longrightarrow}   U(1)  \stackrel{v_{2}}{\longrightarrow}   {\mathbbm 1}\;,  \qquad  v_{1}  \gg   v_{2}.   \label{breakingSU2} 
\eeq
    When the smaller vacuum expectation value (VEV) $v_{2}$ is neglected, i.e., in a high-energy approximation,  the system is known to possess stable regular 't Hooft-Polyakov monopole solutions, with magnetic charge \footnote{As noted by 't Hooft, this result  is consistent with Dirac's quatization condition,    $g_{e} \, g_{m} =  n /2, $  $\, n \in  {\mathbbm Z}$, 
as the smallest charge of the system is that of a quark which can always be introduced in the theory --  with coupling $g_{e}= g / 2$.}, 
\[     g_{m} =    \frac{ n}{g} , \qquad n = 0, \pm 1, \pm 2, \ldots\;.  
\] 
  The $\theta$ term 
\beq    \Delta L =  \theta  \frac{g^{2}}{32 \pi^{2}} \int d^{4}x\,   F_{\mu \nu} {\tilde F}^{\mu \nu} =    
\theta \,  \frac{g^{2}}{8 \pi^{2}}     \int d^{4} x\,        {\bf E} \cdot {\bf B}\:  \label{thetat} \eeq 
induces an electric charge on the magnetic monopole (in units of $g$) of the amount \cite{Witten}
\beq    Q_{e}  =  \frac{\theta}{2 \pi} n  \;. \label{Witten}
\eeq
 Indeed,  the magnetic field of the monopole (seen far from the monopole center), 
\beq   {\bf B}  =  \nabla  \, \frac{ g_{m} }  {r} \;, 
\label{magflux}  \eeq
implies an  electrostatic energy of a pointlike charge, Eq.~(\ref{Witten}).    The latter  can be written also as
\beq    {\bf E} =  \frac{\theta g^{2} }{8 \pi^{2}}   {\bf B }\;,  
 \label{Weffect}\eeq
where both $ {\bf B }$ and  ${\bf E }$ are radially emanating from the monopole center. 

\begin{figure}
\begin{center}
\includegraphics[width=2.8in]{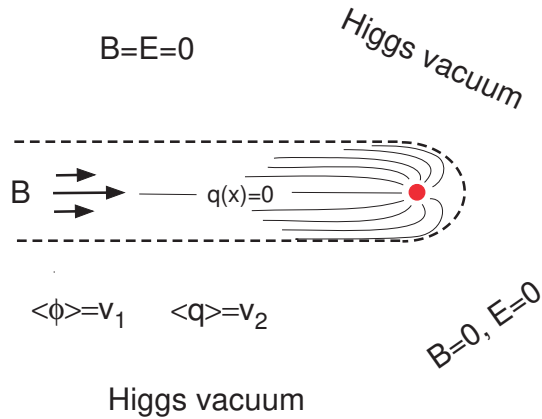}
\caption{A magnetic monopole and the vortex attached to it, immersed in a Higgs vacuum.  Magnetic fields are squeezed to thin flux tubes by the scalar field VEVs.}
\label{MVComplex}
\end{center}
\end{figure}

Even with $v_{2}\ne 0$,  such an approximation should be valid if we look at the system sufficiently close to the monopole, at distance scales between $1/v_{1}$ and $1/ v_{2}$ from the center. When we observe the system from larger distances (larger than $1/v_{2}$),  however, the breaking of the low-energy $U(1)$ group becomes visible, and we will find that the magnetic field ${\bf B}$ is actually squeezed into an Abrikosov-Nielsen-Olesen (ANO) vortex, e.g., in the $-\hat z$  direction (Fig.~\ref{MVComplex}). The system is described at  low energies,  e.g., by the standard Abelian Higgs model with a quartic coupling, 
\beq   {\cal L}=- \frac{1}{4} F_{\mu \nu}^{2} + |{\cal D}_{\mu} q|^{2}-  \lambda \, ( |q(x)|^{2} - v_{2}^{2})^{2},  \qquad  \bra  q \ket =  v_{2}\;. 
\label{AHM}\eeq
The properties of the vortex defect in such a vacuum are well known.

\section{Dual macroscopic description \label{macro}}

Recently a macroscopic description of the $SU(2)$ system with a hierarchical symmetry breaking (\ref{breakingSU2}) has been presented by Chatterjee and Lahiri  \cite{CHA2}.  In analysing the effect of the $\theta$ term, we first follow their work and adopt a dual macroscopic picture of the monopole-vortex complex.  In their approximation, the monopole in Fig.~\ref{MVComplex} is a point and the vortex is a thin line. 
 At $v_{1}$ the VEV of  an adjoint scalar $\phi$  breaks the $SU(2)$ symmetry to $U(1)$; at a much smaller scale $v_{2}$ 
a (scalar) quark field $q(x)$  acquires a VEV breaking the gauge  symmetry completely.  At the mass scales much lower than $v_{1}$,  and in the presence of a monopole,  
\[ \phi(x) =  v_{1}\, {\Phi}(x), \quad  \Phi(x) = U(x) \frac{\tau^{3}}{2}  U^{\dagger}(x)    \]
 with nontrivial winding  $U(x)$, the system is described by a Lagrangian,
\beq    {\cal L}= - \frac{1}{2} \Tr \,  G_{\mu \nu} G^{\mu \nu} +    |{\cal D}_{\mu} q|^{2}    -  \lambda \, ( |q(x)|^{2} - v_{2}^{2})^{2}
\eeq
where   
$    G_{\mu \nu} = F_{\mu\nu} {\Phi}+\hat{M}_{\mu\nu} $,  
and 
\[ F_{\mu\nu}=\partial_{\mu}   A_{\nu}-\partial_{\nu}   A_{\mu} \qquad 
   \hat{M}_{\mu\nu}=\frac{i}{g}   [\partial_{\mu}{\Phi},  \partial_{\nu} {\Phi} ]   \]
are the unbroken $U(1)$ gauge field tensor  and  the field of the magnetic monopole, respectively.  As for the quark, the low-energy degrees of freedom may be taken to be the phase $\chi$ of a $U(1)$ rotation around the $\phi$ direction,  
\beq  q(x) &=&  v_{2}\,  e^{i \chi  \Phi }\, U(x) \,\left(\begin{array}{c}1 \\0\end{array}\right) =  v_{2}\,  e^{i \tfrac{\chi}{2}}\, U(x) \,\left(\begin{array}{c}1 \\0\end{array}\right)   \equiv v_{2}\, e(x),    
\eeq
which leads to  the expression
\beq     |{\cal D}_{\mu} q|^{2} =   \frac{v_{2}^{2}}{4}  \, ( \de_{\mu} \chi - g \, C_{\mu} -  g\, N_{\mu} )^{2},   
\eeq
where $N_{\mu} =   \frac{2 i}{g} {\bar e} \, \de_{\mu}U\, U^{\dagger}  e$  is the gauge field  associated to the monopole and in fact
($M_{\mu \nu}\equiv   2\, \Tr [{\hat M}_{\mu \nu} \, {\Phi}]$)  
\[     M_{\mu \nu} =  \de_{\mu} N_{\nu} - \de_{\nu} N_{\mu}\;.  
\]
The total Lagrangian is then 
\beq     {\cal L}  =  - \frac{1}{4}  (F_{\mu \nu} + M_{\mu \nu})^{2} +  \frac{v_{2}^{2}}{4}  \, ( \de_{\mu} \chi - g \, A_{\mu} -  g\, N_{\mu} )^{2}\;.\label{tothis}
\eeq
In order to dualize \cite{Orland} the fluctuations of the squark field,  we first separate it  into the regular and singular parts $ \chi^{r}$ and  $ \chi^{s}$ 
\beq    \chi =  \chi^{r} + \chi^{s}\;:
\eeq
the latter (non-trivial winding of the scalar field) is related to the vortex position by
 \beq  &&  \epsilon^{\mu \nu \rho\sigma}\de_{\rho} \de_{\sigma} \chi^{s}=  \Sigma^{\mu \nu}(x)   \nonumber \\
 &=& 2\pi n \, \int_{\Sigma} \de_{a}  x^{\mu} \de_{b}x^{\nu}  
(d\xi^{a} \wedge d \xi^{b})\, \delta^{4}(x-x(\xi)) \;
\nonumber   \eeq 
and $\xi^{a}$  are the world-sheet coordinates.  $\chi^{r}$ can be integrated out by introducing the Lagrange multiplier $b_{\mu}$
\beq  - \frac{1}{v_{2}^{2}} b_{\mu}^{2}  +  b_{\mu} \,   ( \de_{\mu} \chi^{r} + \de_{\mu} \chi^{s}   - g \, A_{\mu} -  g\, N_{\mu} )\;,   
\eeq
which gives rise to a functional delta function
\beq   \delta (\de_{\mu}   b_{\mu}(x))\;.
\eeq
The constraint can be solved by introducing antisymmetric fields  $B_{\mu \nu}(x)$, 
\beq    b^{\mu} =  \frac{v_{2}}{2 \sqrt{2}} \,    \epsilon^{\mu \nu \rho \sigma}   \de_{\nu} B_{\rho \sigma} =   \frac{v_{2}}{6 \sqrt{2}}    \epsilon^{\mu \nu \rho \sigma} H_{\nu \rho \sigma }\;, 
\eeq
\beq 
H_{\nu \rho \sigma } \equiv   \de_{\nu} B_{\rho \sigma} +  \de_{\rho} B_{\sigma \nu} +  \de_{\sigma} B_{\nu \rho}  \nonumber
\eeq
  being a completely antisymmetric tensor field.  One is left with the 
Lagrangian 
\beq   - \frac{1}{4}  (F_{\mu \nu} + M_{\mu \nu})^{2}  - \frac{m}{2}  \, 
\epsilon^{\mu \nu \rho \sigma}  A_{\mu} \de_{\nu} B_{\rho \sigma}  + \frac{1}{12}  H_{\mu \nu \lambda}^{2}     +  \frac{m}{ 2 g}   \, B_{\mu \nu} \Sigma^{\mu \nu} - \frac{m}{4}  \epsilon^{\mu \nu \rho \sigma}   M_{\mu \nu}  B_{\rho \sigma}\;
\eeq 
where we have set   
\beq m \equiv \frac{g \, v_{2}}{\sqrt{2}}\, . \eeq 

Now we dualize  $A_{\mu}$ by writing
\bea  &&  \int [d A_{\mu}]    \exp{i \int d^{4}x \,  \{  - \frac{1}{4}  (F_{\mu \nu} + M_{\mu \nu})^{2}  - m\, 
\epsilon^{\mu \nu \rho \sigma}  A_{\mu} \de_{\nu} B_{\rho \sigma} \}} \nonumber \\
&& =    \int \, [d A_{\mu}]  [d \chi_{\mu \nu}]  \,  \exp{i \int d^{4}x \,  \{  -   \chi_{\mu \nu}^2  +     \chi_{\mu \nu} \, \epsilon^{\mu \nu \rho \sigma} (F_{\rho \sigma}  +   M_{\rho \sigma} )/2   - m \,  
\epsilon^{\mu \nu \rho \sigma}  A_{\mu} \de_{\nu} B_{\rho \sigma}/2 \} }  \nonumber \\
&& =   \int    [d \chi_{\mu \nu}]   \,  \delta ( \epsilon^{\mu \nu \rho \sigma}    \de_{\nu}    (\chi_{\rho \sigma}-  {m}    B_{\rho \sigma}/2))  
 \exp{i \int d^{4}x \,  \{  -   \chi_{\mu \nu}^2  +     \chi_{\mu \nu} \epsilon^{\mu \nu \rho \sigma} M_{\rho \sigma} /2 \} }
\label{from}\eea
Again the constraint can be solved by setting 
\beq     \chi_{\mu \nu} =   \frac{1}{2}(  \de_{\mu}   A_{D\,  \nu} - \de_{\nu}   A_{D\,  \mu} +   m\,  B_{\mu \nu} ) 
\label{andfrom}  \eeq
and taking the dual gauge field $A_{D\, \mu}$ as the independent variables. As 
\beq      j^{\mu} =  \de_{\nu} \, \frac{1}{2} \epsilon^{\mu \nu \rho \sigma}  M_{\rho \sigma}  =   \de_{\nu} \, {\tilde   M}^{\mu \nu}  
\eeq 
represents the monopole current,    one sees from Eq.~(\ref{from}) and Eq.~(\ref{andfrom})  that $A_{D}^{\mu}$ is locally coupled to it. 
The Lagrangian is now
\beq  {\cal L}=   \frac{1}{12}  H_{\mu \nu \lambda}^{2}    - \frac{1}{4}(  \de_{\mu}   A_{D\,  \nu} - \de_{\nu}   A_{D\,  \mu}  +    m   \, B_{\mu \nu})^{2}  
+ \frac{m}{2g}  \, B_{\mu \nu} \Sigma^{\mu \nu}  +  A_{D \mu } \, j^{\mu}\;.
\eeq 
Finally, observing that there is a (dual) gauge invariance of the form,  
\beq  \delta  B_{\mu \nu} = \de_{\mu} \Lambda_{\nu} -  \de_{\nu} \Lambda_{\mu};\qquad  \delta A_{D}^{\mu}=  - m\, \Lambda^{\mu}\;,   \eeq
one  may introduce  gauge-invariant fields 
\beq G^{\mu \nu}\equiv    B^{\mu \nu} +   ( \de^{\mu} A_{D}^{\nu} -  \de^{\nu} A_{D}^{\mu})/m \;,   \qquad H_{\nu \rho \sigma } \equiv   \de_{\nu} G_{\rho \sigma} +  \de_{\rho} G_{\sigma \nu} +  \de_{\sigma} G_{\nu \rho}\;,
\eeq
finding the final form of the Lagrangian, 
\beq  {\cal L}=   \frac{1}{12}  H_{\mu \nu \lambda}^{2}    - \frac{m^{2}}{4} \, G_{\mu \nu}^{2} +    \frac{m}{2g} \, G_{\mu \nu}   \Sigma^{\mu \nu}\;.  \label{final}
\eeq
Note however that the integration over $A_{D}^{\mu}$ has introduced a constraint
\beq        \de_{\mu}\, \Sigma^{\mu \nu} = -  g \,  j^{\nu},  \label{constraint}
\eeq
showing that the monopole current acts as the source for the worldsheet fluctuations. In other words,  the monopole is at the endpoint of the vortex (Fig.~\ref{MVComplex}).  
To summarize,  one ended up with a system of massive fields    $G_{\mu \nu}$ only,    coupled to the vortex  world sheet fluctuation ($\Sigma^{\mu \nu}$) and, 
through (\ref{constraint}), to the monopole current ($ j^{\nu}$).
The equations of motion for $G_{\mu \nu}$  gives 
\beq    \de_{\sigma}  H^{\sigma \mu \nu} =  - m^{2} \,   G^{\mu \nu} +  \frac{ m}{g} 
\, \Sigma^{\mu \nu}\;.   \eeq
By taking a further derivative and by using Eq.~(\ref{constraint})  (or by considering the equation of motion of  $A_{D}^{\mu}$ directly)   one finds 
\beq  \de_{\mu}   G^{\mu \nu} = \frac{1}{m}  \,  j^{\nu}\;.  
\eeq

In the presence of a $\theta$ term, Eq.~(\ref{thetat}), one must add to Eq.~(\ref{tothis})
\beq  \Delta  {\cal L}  =   \frac{\theta g^{2}}{32 \pi^{2}}(F_{\mu \nu}+M_{\mu\nu}) ({\tilde F}^{\mu\nu}+{\tilde M}^{\mu\nu})\;.  \label{thetaterm} \eeq
 The electromagnetic duality transformation from Eq.~(\ref{from}) to Eq.~(\ref{final})  gets modified in a standard fashion.
By substituting   ($\alpha \equiv  \theta g^{2} / 8 \pi^{2}$) 
\beq    F_{\mu \nu}^2  \to     F_{\mu \nu} F^{\mu \nu}   -   \alpha F_{\mu \nu} {\tilde F}^{\mu \nu}   =   (a_{+}  F_{\mu \nu} - a_{-} {\tilde F}_{\mu \nu})^{2}   \label{modify1} \eeq
where 
\beq       a_{+}^{2} - a_{-}^{2} =1, \qquad  2\, a_{+} a_{-} = \alpha  =   {\theta g^{2} }/{ 8 \pi^{2}}\;, 
\eeq
Eq.~(\ref{andfrom}) is replaced by 
\beq     a_{+} \chi_{\mu \nu} - a_{-} {\tilde \chi}_{\mu \nu} =    \frac{1}{2}(  \de_{\mu}   A_{D\,  \nu} - \de_{\nu}   A_{D\,  \mu} +   m\,  B_{\mu \nu} ) =   \frac{m}{2}\,  G_{\mu \nu}. 
\eeq  
Solving for $\chi_{\mu \nu}$ this gives 
\beq  \chi_{\mu \nu}=    \frac{m}{2}   \frac{1}{a_{+}^{2} + a_{-}^{2}}  ( a_{+}   G_{\mu \nu} + a_{-}   {\tilde G}_{\mu \nu} ) \;;
\eeq
 the final form of the Lagrangian is 
\beq  {\cal L}=   \frac{1}{12}  H_{\mu \nu \lambda}^{2}    - \frac{m^{2}}{4 \, ( 1+\alpha^{2} ) }  ( G_{\mu\nu}  G^{\mu\nu} + \alpha  \, G_{\mu\nu}\tilde{G}^{\mu\nu})   +    \frac{m}{2 g} \,  G_{\mu \nu}   \Sigma^{\mu \nu}\;,  \label{finalBis}
\eeq
instead of Eq.~(\ref{final}).    Actually,  we must write  $F_{\mu \nu} +M_{\mu \nu}$ instead of $F_{\mu \nu}$ above, but the way  $M_{\mu \nu}$ enters the final result through the constraint Eq.~(\ref{constraint}) is not modified.

The equations of motion  following from Eq.~(\ref{finalBis}) are:  
\beq \partial_{\sigma}H^{\sigma\mu\nu}=-\frac{m^{2}}{1+\alpha^{2}} (G^{\mu\nu}+\alpha \tilde{G}^{\mu\nu})   +\frac{m}{g}\Sigma^{\mu\nu},   \label{EqMotfinal}   \eeq
\beq   \frac{1}{1+\alpha^{2}}  \partial_{\mu} (G^{\mu\nu}+  \alpha \tilde{G}^{\mu\nu})= - \frac{1}{m}   j^{\nu}\label{theta2}\;, \qquad \alpha \equiv  \frac{\theta g^{2}}{ 8 \pi^{2}}  \;.\label{eqmotMod}
\eeq
    To solve Eqs.~(\ref{EqMotfinal})  and  (\ref{eqmotMod}),    set  
 \beq  K_{\mu}\equiv      \de^{\lambda} G_{\mu  \lambda}, \qquad L_{\mu}\equiv    \de^{\lambda} {\tilde G}_{\mu  \lambda}=\frac{1}{6} \epsilon_{\mu \nu \rho \sigma} H^{\nu \rho \sigma}\;.
 \eeq
 One has from Eq.~(\ref{eqmotMod})  
 \beq     \frac{1}{1+\alpha^{2}}   (K_{\mu} + \alpha L_{\mu} )  =  \frac{1}{m} \,  j_{\mu}\;, \label{combine1}
 \eeq
 while from  Eqs.~(\ref{EqMotfinal}) 
 \beq       -  (\de_{\mu}  L_{\nu} - \de_{\nu} L_{\mu})  +\frac{m^{2}}{1 + \alpha^{2}}    ({\tilde G}_{\mu \nu} - \alpha G_{\mu \nu})  = \frac{m}{g}  {\tilde \Sigma}^{\mu\nu},\label{reduces} 
 \eeq
 and hence
 \beq      \de^{\mu} \de_{\mu}  L_{\nu}  + \frac{m^{2}}{1 + \alpha^{2}}   (L_{\nu} - \alpha K_{\nu}) =     m  \,  {\tilde j}_{\nu}\;,   \label{combine2}
 \eeq
 where we have defined 
 \beq       {\tilde j}^{\nu} \equiv   -\frac{1}{g}      \de_{\mu}\, {\tilde \Sigma}^{\mu \nu} , \qquad   {\tilde \Sigma}^{\mu \nu}  \equiv \frac{1}{2}  \epsilon^{\mu \nu \rho \sigma} \Sigma_{\rho \sigma}\;.
 \eeq
 Combining Eq.~(\ref{combine1}) and  Eq.~(\ref{combine2})  we have an explicit solution for $L_{\mu}$:
 \beq     \de^{\mu} \de_{\mu}  L_{\nu}  + m^{2}  L_{\nu }  =  m  \, (\alpha  j_{\nu} + {\tilde j}_{\nu} )\;, \qquad .^{.}.  \quad L_{\mu}=  \frac{m}{\Box +m^{2}}    \, (\alpha  j_{\nu} + {\tilde j}_{\nu} )\;.  \label{result}
 \eeq
 In order to interpret the result in terms of the original electric and magnetic fields, we note that the duality transformation Eq.~(\ref{modify1})-Eq.~(\ref{finalBis})   implies 
\bea   F_{\mu \nu} &=&  -\frac{m}{1 + \alpha^{2}}    ({\tilde G}_{\mu \nu} - \alpha G_{\mu \nu})   =  - \frac{1}{g} {\tilde \Sigma}_{\mu \nu}  -  \frac{1}{m}  (\de_{\mu}  L_{\nu} - \de_{\nu}  L_{\mu}) \;  \nonumber \\
&=&     - \frac{1}{g} {\tilde \Sigma}_{\mu \nu}  -      \frac{1}{\Box +m^{2}}    \left[  \de_{\mu} (\alpha  j_{\nu} + {\tilde j}_{\nu} )  -   (\mu \leftrightarrow  \nu)      \right]\;.
 \label{elmagfield} \eea

For instance, let us consider a massive static monopole sitting at ${\bf r}=0$ with a vortex attached to it and extending into the $-{\hat z}$ direction:  
\beq   \Sigma^{30}=-\Sigma^{03} =4 \pi \, n\, \delta(x)\delta(y) \theta(-z)\;, \qquad    \Sigma^{\mu \nu}=0 \quad (\mu \nu) \ne (30), (03)\;; \eeq 
\beq   j^{0}  =   \frac{ 4 \pi \, n}{g}\, \delta^{3}({\bf r}),  \quad  j^{i}=0\;; \quad i=1,2,3\;;  \qquad   {\tilde j}^{\nu} =  -\frac{1}{g}      \epsilon^{\lambda  \nu  0  3}  \, \de_{\lambda} \Sigma_{03}\;.
\eeq
 From Eq.~(\ref{elmagfield}) one finds that (we recall $\alpha= \theta g^{2}/  8 \pi^{2}$)
 \beq     E_{i} =  F_{0i}=   \alpha\, B_{i}^{(mon)}, \qquad B_{i} = \frac{1}{2} \epsilon_{ijk} F_{jk}=     B_{i}^{(mon)} +   B^{(vor)}   \delta_{i}^{3}\;,  \label{remark}
 \eeq
 where 
 \beq      B_{i}^{(mon)} =  \frac{n}{g}  \de_{i}  G({\bf r}), \qquad  B^{(vor)}= \frac{n}{g} \,   m^{2} \,  \int_{-\infty}^{0} \, dz^{\prime} \,  G(x, y, z-z^{\prime})\;,\label{using}
 \eeq
 and $G({\bf r})$ is the Green function, having the Yukawa form
 \beq     G({\bf r}) = \frac{4\pi}{- \Delta + m^{2}}\, \delta^{3}({\bf r}) =     \frac{e^{-m r}}{r}\;.  \label{using2}
 \eeq
   Note the clear-cut separation of the monopole and vortex contributions to magnetic (and electric) fields,  Eq.~(\ref{remark}).

   In order to see magnetic Gauss' theorem at work, let us integrate the magnetic flux through the surface of a sphere centered at the origin (the monopole position), of an arbitrary radius $R$, 
   \beq        \Phi (R)  =   \int_{\de S} d{\bf S}\cdot  {\bf B}\;,   
   \eeq
   that is, 
   \beq    \Phi (R)  =     \int_{S} d^3r   \,  \de_{i} (  B_{i}^{(mon)}   +  B^{(vor)} \delta_{i}^{3} )=   \frac{n}{g} \,\int_{S} d^3r   \,  \Delta   G({\bf r}) + 
  \int_{S} d^3r   \,   \de_{3}  B^{(vor)} \;.
   \eeq
   By using Eq.~(\ref{using}) and Eq.~(\ref{using2})  we see that
   \beq   \Phi (R) =   -  \frac{4 \pi n}{g}\;,
   \eeq
   independently of the radius $R$.  In other words,    \[  \de_{i}  B_{i} =  - \frac{4\pi n }{g} \delta^{3}({\bf r})\;.  \] 
   In the limit of  very small ($\ll 1/m$) and large  ($\gg 1/m$) values of $R$,  the above result reduces to  the ``flux-matching''  condition between the magnetic monopole and vortex flux. Namely,  
  \beq        \int_{R\ll 1/m} dS \cdot {\bf B}^{(mon)} \simeq  \Phi(R)|_{R \ll 1/m} =   \Phi(R)|_{R \gg 1/m}  \simeq  - \int_{|z|\gg 1/m}  dx\, dy\,      B^{(vor)}  \;,  \eeq
  as can be verified explicitly. 
 
 Electric field takes significantly nonzero values   only near the monopole: Gauss'  theorem in the usual form  does not hold because the electric charge of the monopole is screened by the charge condensed in the vacuum.  In the small spherical region of Coulomb vacuum surrounding the monopole it is proportional to magnetic field, consistently with Witten's formula, Eq.~(\ref{Weffect}).   
 
 \section{Microscopic description}  
 
It might be of some interest to know how the electric field behaves near the vortex-monopole complex (the behavior of the magnetic field in an ANO vortex is well known).
 In order to have a microscopic description of the monopole-vortex complex let us take as the model an $SU(2)$ gauge theory with  softly broken ${\cal N}=2$  supersymmetry.
 One of the motivations for doing so, rather than sticking to the simple model of Section~\ref{macro},   
  is that the generalization to the non-Abelian vortex case is straightforward in such a context:  it is in fact in such a context that the vortex solutions with non-Abelian moduli have been found.  Also, the form of the potential is not modified by radiative corrections due to nonrenormalization theorem of supersymmetric theories.  Finally, the interesting phenomenon of orientational zero modes and their fluctuations in the case of non-Abelian vortices (see below) have so far been studied only in such a microscopic picture.  
 
 The Lagrangian is of  the form,  
\beq
{\cal L}=     \frac{1}{ 8 \pi} \im \, S_{cl} \left[\int d^4 \theta \,
\Phi^{\dagger} e^V \Phi +\int d^2 \theta\,\frac{1}{ 2} W W\right]   + {\cal L}^{(quarks)}  +  \int \, d^2 \theta \,\mu  \,\Tr  \Phi^2;  
\label{lagrangian}
\eeq
\beq {\cal L}^{(quarks)}= \sum_i \, [ \int d^4 \theta \, \{ Q_i^{\dagger} e^V
Q_i + {\tilde Q_i}  e^{-V} {\tilde Q}_i^{\dagger} \}   +  \int d^2 \theta
\, \{ \sqrt{2} {\tilde Q}_i \Phi Q^i    +      m_{0}\,   {\tilde Q}_i Q^i   \}
\label{lagquark}
\eeq
where $m_{0}$ is the bare quark masses, and where 
\beq
S_{cl} \equiv  \frac{\theta_0 }{ \pi} + \frac{8 \pi i }{ g_0^2}.  
\label{struc}
\eeq   
The parameter $\mu$,  the mass of the adjoint chiral multiplet which breaks the supersymmetry to ${\cal N}=1$, is taken to be small as compared to the
bare quark mass $m_{0}$. The adjoint scalar takes a VEV, $\langle \phi \rangle  =  v_{1} \diag (1, -1)$,  where  $v_{1}= - m_{0}/\sqrt{2}$,  which breaks the gauge symmetry to $U(1)$.     The upper component of the squark remains massless  and at much lower energies its VEV  $v_{2}= \langle q \rangle =  2  \sqrt{\mu m_{0}}$  breaks 
the $U(1)$ symmetry.  In other words,  the mass parameters  are chosen as 
\[     |m_{0}| \gg  |\mu|,    \qquad     v_{1} \gg v_{2}\;:
\]
so that the gauge symmetry is broken at two, hierarchically different scales.   The model considered here is actually identical to the one analyzed in some detail earlier 
\cite{ABEK,AEV} apart from the simplification to $SU(2)$ theory, Eq.~(\ref{breakingSU2}),  rather than  $SU(3)$ case studied there.  
The low-energy bosonic Lagrangian takes the form   ($\phi = v_{1}\, {\rm diag} (1, -1)  + \lambda(x)   $) 
\bea  {\cal L}   &=&  \int d^{4}x  \Big[\, 
 -  \frac{1}{4 g^{2}}  \, ( F_{ij}^{0})^{2} +   \frac{1}{2 g^{2}}  (F_{0i}^{0})^{2} +
  \frac{1}{ g^{2}}  |\de_{\mu}  \lambda^{3}|^{2}  \nonumber  \\
  &&  +    | {\cal D}_{\mu} q|^{2}  
  -   g^{2} | \mu \phi^{3}  +  \sqrt{2} Q^{\dagger} t^{3} Q   |^{2}  
  -    2 \, Q^{\dagger}  \lambda^{\dagger} \lambda  Q    +       \frac{\theta}{8 \pi^{2}} F_{0i}    {\bf B}_{i}  \Big]    
 \label{ffrom} \eea
  The presence of $\mu  \, \phi_{3}\propto  \mu \,m_{0}$ further breaks $U(1)$ completely and gives rise to vortex.  Since ($t^{3}=\tau^{3}/2$)
\[    {\cal D}_{\mu} Q = (\de_{\mu} -   i \, t^{3} A_{\mu}^{3} ) Q,
\]
the light quark (the upper component of $Q$) enters with the covariant derivative as
\[  {\cal D}_{\mu} \, q =   \left( \de_{\mu} -  i  {A_{\mu}^{3}}/{2} \right) \, q\;.    
\]
In order to study the monopole-vortex complex configuration it is necessary to work in the so-called singular gauge, in which all fields smoothly approach their constant VEVs away from the monopole-vortex region, without any  ``winding''. 
The gauge field $A_{\phi}$ presents a Dirac string singularity
\beq    A_{\phi}^{3} \sim   - \frac {2 }  {\rho}\;,   \quad z <0\;, \qquad  \rho \equiv \sqrt{x^{2}+ y^{2}},  \label{Dirac}
\eeq
 along the vortex core in such a gauge,  but the (light) squark field vanishes there,  making it  innocuous. 

We work with an  Ansatz for the form of the fields far from the monopole center (the suffix $3$ referring to the third direction in the $SU(2)$)
\beq    A_{\phi}^{3} =  -    \frac{2}{\rho}  f(\rho, z)    , \qquad   A_{0}^{3}=   A^{0}(\rho, z)   \;, \qquad   q(x) = w(\rho, z) \;;
\label{AAnsatzebis}\eeq
\[ \phi ({\bf r})  =   \left(\begin{array}{cc}  v_{1} & 0 \\0 & -v_{1} \end{array}\right) + \lambda(\rho, z) , \qquad 
 \lambda (\rho, z) =  t^{3} \lambda_{3}(\rho, z) \;,   \]
where  cylindrical coordinates $\rho, \phi, z$ are used.  The four profile functions $f, A^{0}, w, \lambda$ satisfy coupled second-order differential equations and we must impose an appropriate set of  boundary conditions so that the configuration approaches
't Hooft-Polyakov's radial solution near the monopole, and a vortex-type solution far from it.  The details will be presented elsewhere \cite{Cinque}, 
together with a numerical solution of the problem.
  
The behavior of magnetic field around the vortex far from the monopole is determined by the standard boundary condition 
\beq    f(\rho, z) \to f(\rho),  \qquad  w(\rho, z) \to w(\rho)\;, 
\eeq
\beq    f(0) = 1; \quad f(\infty)=0, \qquad   w(0)=0, \quad w(\infty)=1\;.
\eeq
  It is not modified by the $\theta$ term, in agreement with what was found in the macroscopic picture.  It  approaches a constant  at the vortex core and exponentially suppressed  as $K_{0}(\rho /\sqrt{2} )$ at large $\rho$ (where we have set $g  = v_{2}=1$), where  $K_{0}(x)$ is a modified Bessel function of the second kind.  The total magnetic flux carried by the vortex, measured far from the monopole, is given by
\beq     \int d^{2} x  B_{z} =   -  2\pi \int_{0}^{\infty}   d\rho\, \rho  \frac{2}{\rho} 
\frac{\de f(\rho)}{\de \rho}  =   + \frac{4\pi}{g}\;. 
\label{correctly}\eeq
On the other hand,  near the monopole center, i.e., at distances much less than $1/ v_{2}$,  the field configuration is well approximated by the 't Hooft-Polyakov solution: it is not affected significantly by the smaller VEV, $v_{2}$.  Its total magnetic flux $-4\pi /g $ through the surface of a sphere surrounding it (following from Eq.~(\ref{magflux})) matches correctly the vortex flux Eq.~(\ref{correctly}).
%matches to the monopole flux  correctly.  

The behavior of electric field is slightly subtler. The equation of motion for $A^{0}$  following from Eq.~(\ref{ffrom}) is:
 \beq    \Delta   A^{0}  -  m^{2}   A^{0}  \, w^{2}  =   \frac{\theta}{2\pi}  \delta^{3}({\bf r}) \;, \qquad  m\equiv   \frac{g v_{2}}{\sqrt{2}}\;, 
\label{EFieldsimp}\eeq
where on the right hand side we have made an approximation for $\nabla \cdot {\bf B}$ which is valid far from the monopole center.
Away from the region of monopole {\it and} vortex, 
\[ w = |q|  \to   1, \]
  and we find the regular solution 
\beq     A^{0} \simeq    - \frac{\theta g^{2}}{8 \pi^{2}} \frac{e^{- m r}}{r}= - \frac{\theta g^{2}}{8 \pi^{2}}  G({\bf r})  \;, 
\qquad  {\bf E} = - \nabla A^{0},  \label{Regsolut} \eeq
consistently with Eq.~(\ref{remark}).
Such an asymptotic behavior is valid in all directions except near the negative $z$ axis (i.e., along the vortex), where it is distorted by the fact that $q=w$ drops to zero as
$w \sim \rho$    ($\rho= \sqrt{x^{2}+y^{2}}$ is  the distance from the vortex axis).

It might be thought that electric field survives in the vortex core, as does magnetic field. A constant ($z$-independent)  electric field is however not consistent with  
Eq.~(\ref{EFieldsimp}), which reads  in cylindrical coordinates,   
\beq    \frac{\de^{2}A^{0} }{\de \rho^{2}} +    \frac{1}{\rho}   \frac{\de A^{0} }{\de \rho}  +   \frac{\de^{2} A^{0} }{\de z^{2}}  -  m^{2}   A^{0}  \, w^{2}  =  \frac{\theta}{2\pi}  \delta^{3}({\bf r}) \;.
\label{EFieldcyl} 
\eeq
In the absence of the third term, the equation reduces to 
\beq    \frac{\de^{2}A^{0} }{\de \rho^{2}} +    \frac{1}{\rho}   \frac{\de A^{0} }{\de \rho} -  m^{2}   A^{0}  \, w^{2}  =0\;, 
\eeq
whose solution, suppressed at large $\rho$,  is necessarily singular at $\rho=0$.  Such a behavior is not accepatable as
it leads to singular electric field \footnote{This is in contrast to the unphysical Dirac singularity of $A_{\phi}$, Eq.~(\ref{Dirac}):
 magnetic field is perfectly regular along the vortex core.}.

\begin{figure}
\begin{center}
\includegraphics[width=2.8 in]{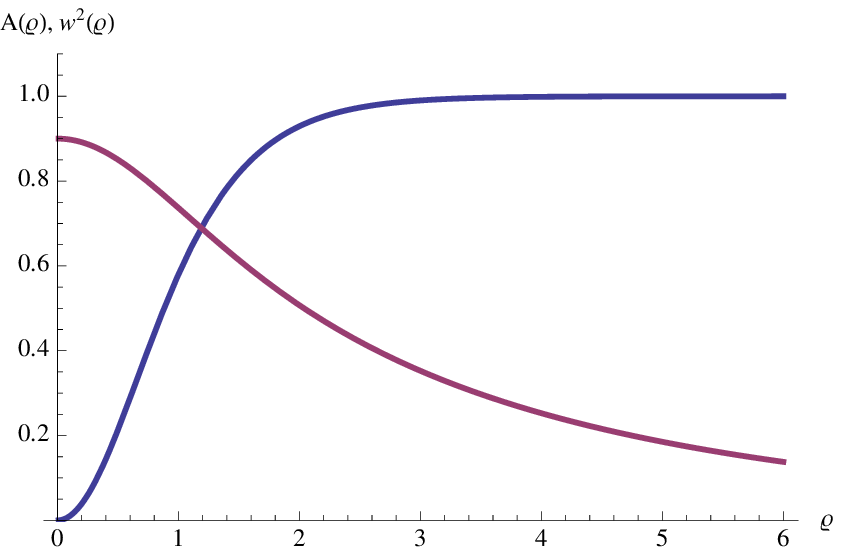}
\caption{ }
\label{Schroed}
\end{center}
\end{figure}

Assuming that the electric potential falls exponentially in $|z|$ and assuming a $\rho$ independent exponent, 
\beq  A^{(0)} \sim  e^{\kappa   z}  {\tilde A}, \qquad    \rho \sim O\left(\frac{1}{m}\right)  \ll   |z|\;,    \label{elepot}  \eeq   
the equation for $ {\tilde A}$ reads 
\beq       - {\tilde A}_{\rho \rho}  -   \frac{1}{\rho}   {\tilde A}_{\rho}      +  m^{2}  \, w^{2}     {\tilde  A} =\kappa^{2}    {\tilde A}\;,   \label{Sch}
\eeq
 This has the form of a two-dimensional (radial)  Schr\"odinger equation with ``potential''   $ \tfrac{1}{2}  m^{2}  \, w^{2}$  and  energy $ \tfrac{1}{2} \kappa^{2}$.
 A regular electric potential damped at $\rho=\infty$ corresponds to a bound-state wave function.
We assume that the squark field $ |q| =  w(\rho)$ can well be approximated, at large and negative  $z$, by 
the standard ANO vortex profile function. The latter behaves as  $\propto  \rho$  at small $\rho$ and approaches a constant value $1$ as $\rho \to \infty$, see Fig.~\ref{Schroed}.   With this ``potential''  it can be shown that there is a unique bound state with  $\kappa\simeq 0.93 m$.
 The ``wave function'' looks like the one in Fig.~\ref{Schroed}.  
Electric potential (\ref{elepot}) is exponentially suppressed both in  $|z|$  and in  $\rho$. 
The behavior (\ref{elepot})  at small $\rho$  must be smoothly connected to the region of larger $\rho$,  
\beq   A^{(0)} \sim    \frac{e^{m z}}{|z|} \, e^{m \rho^{2}/ 2 z},  \qquad    \frac{1}{m} \ll  \rho   \ll |z| 
\label{zby}\eeq
which follows from (\ref{Regsolut}), but a better understanding of such a transition would require a careful numerical study of the coupled equations.  We note that the fact that the suppression of electric field along the vortex core ($e^{- 0.93 m |z|}$) is less severe than in the region outside  the vortex-monopole complex ($e^{- m r}$) is physically quite reasonable.  

The correction to the energy   due to  electric field can be estimated by considering 
 a large sphere surrounding the monopole and part of the vortex, with its center at the monopole position, with  radius  $R= |z|$  much larger than the vortex width  ($1/v_{2}$), and studying the energy contained in it due to the terms involving  $A^{(0)}$.  
   The terms containing  $A^{(0)}$  in the energy  are (we set $g=1$ below)
  \bea &&  S  + \int  d^{3}{x } \, \left[    - \frac{1}{2}  A_{0} \Delta A_{0} +  \frac{1}{4}  A_{0}^{2}  w^{2}  -  \frac{\theta}{8 \pi^{2}}   (\nabla \cdot {\bf B})  \, A_{0} \right]\:;  \nonumber \\
 && S=  \int  d^{3}{x } \,    \nabla  \cdot \left[ \frac{1}{2} A_{0}\nabla A_{0}  +  \frac{\theta}{8\pi^{2}} A_{0}{\bf B}\right]\;:
\label{bothterm}  \eea
 $S$ is the surface term.
  
\begin{figure}[h]
\begin{center}
\includegraphics[width=2.5 in]{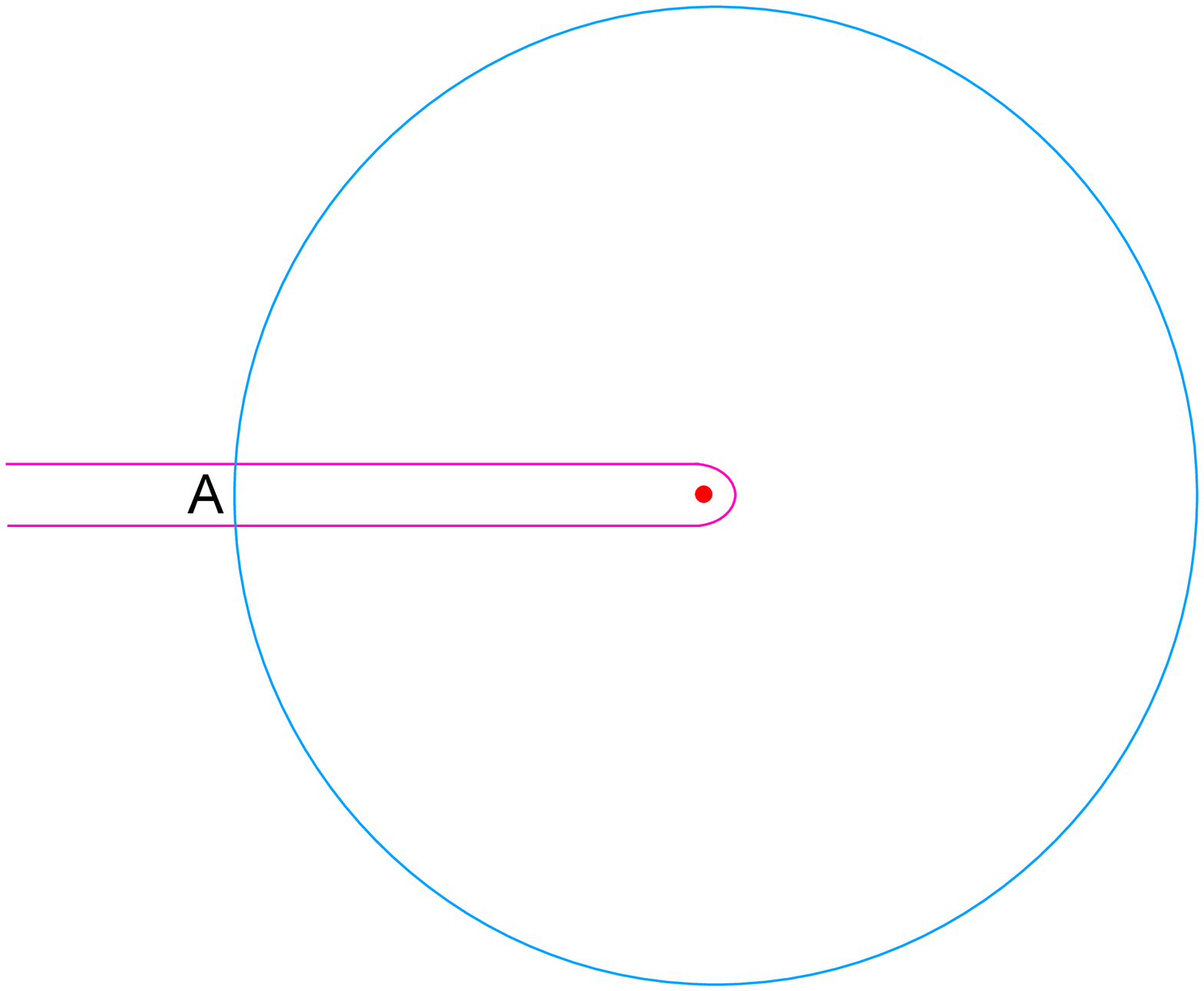}
\caption{}
\label{Energy}
\end{center}
\end{figure}

  The bulk term gets a nonvanishing contribution 
 from a small region around the monopole (${\bf r}=0$), as $A^{(0)}$ obeys the equation of motion ~(\ref{EFieldsimp}) (see Eq.~(\ref{magflux})).  It is a 
correction to the monopole mass   proportional to  $\theta^{2}$. 
   Contribution to  the surface term $S$  vanishes  almost everywhere in the Higgs vacuum, except possibly from a small surface region where the vortex cuts through the sphere (indicated by {\tt A}  in Fig.~\ref{Energy}), where  
 \[ d{\bf S} =  - {\hat z} \, d^{2}x. \]  
 However, as $A_{0}$ is exponentially suppressed far from the monopole  in all directions,  including along the vortex  core,   no surface term arises. 
 The vortex tension is unmodified by the $\theta$ term.

 \bigskip

\section{Outlook} 

We have thus initiated the investigation of a monopole-vortex complex appearing as a result of a hierarchical gauge symmetry breaking, 
focusing our attention to the simplest such system,  
\beq
SU(2) \stackrel{v_{1}}{\longrightarrow}   U(1)  \stackrel{v_{2}}{\longrightarrow}   {\mathbbm 1}\;,  \qquad  v_{1}  \gg   v_{2}\;,     \label{breakingSU2Bis} 
\eeq
and  paying a particular attention to the effects related to the $\theta$ parameter of the underlying  $SU(2)$ theory.   We have found a consistent picture of the 
 $\theta$ dependence  both in a macroscopic (generalized-London-limit-type) approximation and in a microscopic electric description of the vortex-monopole complex soliton.   

 Of course,  our main interest lies in  more nontrivial non-Abelian symmetry breaking scenarios, such as  ($N \ge 2$)
\beq
SU(N+1) \stackrel{v_{1}}{\longrightarrow}  \frac{SU(N) \times U(1)}  {{\mathbbm Z}_{N}}   \stackrel{v_{2}}{\longrightarrow}   {\mathbbm 1}\;,  \qquad  v_{1}  \gg   v_{2} \;, \label{SUNhierarchy} 
\eeq
 as mentioned in the Introduction.     An extensive study of the low-energy vortex solutions possessing non-Abelian continuous orientational moduli  has been performed in the last several years  (for reviews, see \cite{ModMat,Tong,SYBook,Konishi})
after the explicit construction of the {\it non-Abelian vortex} solutions \cite{HT},\cite{ABEKY}.   The model considered is a natural generalization of the Abelian Higgs model, Eq.~(\ref{AHM}),   in which the gauge group is taken to be, e.g., 
\beq   H= SU(N) \times U(1)/{\mathbbm Z}_{N}\sim U(N)\;, \label{nonAbbr}\eeq  
with $N_{f}=N$ squarks in the fundamental representation of $SU(N)$. The ground state of the system is characterized by the scalar quark VEV, 
\beq    \bra q \ket =    v_{2}\, {\mathbbm 1}_{N\times N}\;,  \eeq
where the squark field is written in a color (vertical)-flavor (horizontal) mixed matrix form. 
The system is in the so-called color-flavor locked phase:  the gauge symmetry is broken completely; at the same time the color-flavor diagonal symmetry $SU(N)_{C+F}$  is left unbroken.  A minimum vortex configuration in this vacuum winds the minimum $U(1) \subset  H$, and accordingly
has the form, 
{\small 
\beq   q(x) =   v_{2}\,  \left(\begin{array}{cccc}e^{i \phi} \, f(\rho) & 0 & 0 & 0 \\0 & g(\rho) & 0 & 0 \\0 & 0 & \ddots & 0 \\0 & 0 & 0 & g(\rho)\end{array}\right)\;, \qquad   f(\infty)=g(\infty) = 1,   \eeq  
}
and we considered a particular solution in which the first flavor winds nontrivially.  Such a solution breaks the exact color-flavor symmetry of the system as
\beq     SU(N) \to    \frac{SU(N-1) \times U(1)}  {{\mathbbm Z}_{N-1}} \label{broken}
\eeq
and therefore develops  orientational zeromodes living on  
\beq    CP^{N-1} =    \frac{  SU(N)  }{   SU(N-1)\times U(1)  / {\mathbbm Z}_{N-1} }  \;.  \label{interesting}
\eeq
These correspond to  ``Nambu-Goldstone modes'' freely propagating  however only along the vortex; in the bulk these are massive modes.
In other words,  these modes fluctuate in the vortex worldsheet, and can be described by an effective $1+1$ dimensional $CP^{N-1} $ sigma model \cite{ABEKY,SY, GJK}, 
\beq  S= \frac{4\pi}{g_{N}^{2}} \int \,dz \,dt\,  \left[   \de_{\alpha}  n^{\dagger} \de_{\alpha} n +    ( n^{\dagger}\de_{\alpha} n )^{2}  \right]  \;, \qquad \alpha=z, t,
\eeq
where the complex $N$-component unit vector $n$ represents the coordinates of $CP^{N-1}$.  

Now consider our $H$ system as a low-energy approximation of the underlying $SU(N+1)$  gauge theory, as in (\ref{SUNhierarchy}).  The $\theta$ term of the $SU(N)$ gauge system 
inherited from the original $SU(N+1)$ theory  induces a nontrivial $\theta$-dependent term in the low-energy $CP^{N-1}$ sigma model,
\beq    \Delta S =    \frac{4\pi}{g_{N}^{2}} \int \,dz \,dt\,  \left[   - \frac{\theta g_{N}^{2} }{8  \pi^{2}}  \,  \epsilon_{\alpha \beta}   \,  \de_{\alpha}  n^{\dagger}  \de_{\beta} n   \right]  \;   
\eeq
 as shown by Shifman and others \cite{SY}.  We note that these effects, being magnetic, survive at long distances along the vortex through the fluctuating zeromodes, in contrast to the effects related to the $U(1)$ Witten (electric) charge studied above, concentrated near the monopole.  
 
  It would be very interesting to ask what these $\theta$ dependent magnetic features of the vortex imply for  the property of the massive monopoles sitting at the extremes, in the monopole-vortex complex soliton generated as a result of the hierarchical symmetry breaking, (\ref{SUNhierarchy}).   To answer this question, however,  requires a proper understanding the behavior of the  fluctuating non-Abelian orientational zeromodes {\it of the whole monopole-vortex complex},  a problem currently under investigation. We shall come back to the important issue of the matching of the $\theta$-related effects in the context of non-Abelian monopole-vortex complex  in a separate work  \cite{Cinque} in near future.

 \section*{Acknowledgments}  The authors thank M. Cipriani, D. Dorigoni, J. Evslin, T. Fujimori  and S. B. Gudnason for useful
 discussions.


\begin{thebibliography}{100}

   \bibitem{ABEKY}
R.~Auzzi, S.~Bolognesi, J.~Evslin, K.~Konishi and A.~Yung,
%``Nonabelian superconductors: Vortices and confinement in N = 2 SQCD,''
Nucl.\ Phys.\ B {\bf 673} (2003) 187
[arXiv:hep-th/0307287].
%%CITATION = HEP-TH 0307287;%%
 \bibitem{ABEK}
  R.~Auzzi, S.~Bolognesi, J.~Evslin and K.~Konishi,
  %``Nonabelian monopoles and the vortices that confine them,''
  Nucl.\ Phys.\  B {\bf 686} (2004) 119
  [arXiv:hep-th/0312233];  
 M.A.C.~Kneipp, 
%  Color superconductivity, Z(N) flux tubes and monopole confinement in deformed N=2* superYang-Mills theories.
Phys. Rev. D {\bf 69}: 045007 (2004) 
[arXiv:hep-th/0308086].
  
   \bibitem{Duality}
 M.~Eto, L.~Ferretti,  K.~Konishi, G.~Marmorini, M.~Nitta, K.~Ohashi, W.~Vinci and N.~Yokoi,
%  ``Non-abelian duality  from  vortex moduli: 
%a dual model of  color-confinement'',   
Nucl. Phys. B  {\bf 780} 161-187, 2007
   [arXiv:hep-th/0611313].
   
  \bibitem{Konishi} K.~Konishi, in Lecture Notes in Physics, {\bf 737}  471 (2008),  Springer  
   [arXiv:hep-th/0702102].

\bibitem{TH}  
  G.~'t Hooft,   Nucl. Phys.   B {\bf 79},   817    (1974),  A.M.~Polyakov,  JETP Lett.  {\bf 20},  194   (1974).


\bibitem{CHA2} C.~Chatterjee and A.~Lahiri,  JHEP 1002:033,2010  [arXiv:0912.2168 [hep-th]];    JHEP 0909:010, 2009. 
[arXiv:0906.4961 [hep-th]];   Europhys. Lett. \textbf{76} (2006) 1068,     [arXiv:hep-ph/0605107].


\bibitem{HT}
A.~Hanany and D.~Tong,
%``Vortices, instantons and branes,''
JHEP {\bf 0307}, 037 (2003)
[arXiv:hep-th/0306150].

  \bibitem{SY}
% M. Shifman and A. Yung, Phys. Rev. D {\bf 66},  045012  (2002);    
  M.~Shifman and A.~Yung,  Phys. Rev.  D {\bf 70},  045004  (2004)  [arXiv:hep-th/0403149];  
A.~Gorsky, M.~Shifman and A.~Yung,  Phys. Rev. D  {\bf 71}, 045010 (2005) [arXiv:hep-th/0412082].
  
\bibitem{ModMat}
 M.~Eto, Y.~Isozumi, M.~Nitta, K.~Ohashi and N.~Sakai,
 %``Solitons in the Higgs phase: The moduli matrix approach,''
 J.\ Phys.\ A  {\bf 39}, R315 (2006)
 [arXiv:hep-th/0602170]. 

\bibitem{Tong}
D.~Tong, ``TASI lectures on solitons: Instantons, monopoles, vortices and kinks''
[arXiv:hep-th/0509216],   
  ``Quantum Vortex Strings: A Review,''
  [arXiv:0809.5060 [hep-th]].

\bibitem{SYBook}  M.~Shifman and A.~Yung,
   Rev.\ Mod.\ Phys.\  {\bf 79}, 1139 (2007)
  [arXiv:hep-th/0703267].

\bibitem{GJK} S.B.~Gudnason, Y.~Jiang and  K.~Konishi,   JHEP {\bf 1008}:012 (2010) 
[arXiv:1007.2116 [hep-th]].
  
 \bibitem{APS}   P.C.~Argyres, M.R.~Plesser and N.~Seiberg, Nucl. Phys.  B {\bf 471}, 159  
(1996) [arXiv:hep-th/9603042];   P.C.~Argyres, M.R.~Plesser and  A.D.~Shapere, 
Nucl. Phys. B {\bf 483}, 172 (1997) [arXiv:hep-th/9608129];
 K.~Hori, H.~Ooguri and Y.~Oz,
 Adv. Theor. Math. Phys.   {\bf 1}, 1  (1998) [arXiv:hep-th/9706082].


\bibitem{CKM}
G.~Carlino, K.~Konishi and   H.~Murayama,   
 Nucl. Phys.   B {\bf 590},  37 (2000) [arXiv:hep-th/0005076]. 


\bibitem{Witten} E.~Witten, Phys. Lett. \textbf{86B}, 283  (1979).

\bibitem{Orland}  P.~Orland, Nucl. Phys. B  {\bf  428}, 221  (1994) [arXiv:hep-th/9404140].

  \bibitem{Cinque}  M.~Cipriani, D.~Dorigoni, B.~Gudnason,  K.~Konishi and  A.~Michelini, in preparation.  

\bibitem{AEV}    R.~Auzzi, M.~Eto and W.~Vinci, 
 JHEP {\bf 0711}:090 (2007)  [arXiv:0709.1910 [hep-th]].

\end{thebibliography}
 \end{document}